**Antiferromagnetic Quantum Criticality in Infinite-Layer Cuprates $Sr_{1-x}Nd_xCuO_2$**

Bin-Jie Wu[1], Ze-Xian Deng[1], Guang-Ran Wei[1], Xi Zhou[1], Ji-Hai Zhang[1], Da-Zhuang Deng[1], Qun Zhu[1], Yong Zhong[2*], Can-Li Song[1,3*], Xu-Cun Ma[1,3*], Qi-Kun Xue[1,3,4,5]

[1]*Department of Physics and State Key Laboratory of Low-Dimensional Quantum Physics, Tsinghua University, Beijing 100084, China*

[2]*Quantum Science Center of Guangdong-Hong Kong-Macao Greater Bay Area, Shenzhen, Guangdong Province 518045, China*

[3]*Frontier Science Center for Quantum Information, Beijing 100084, China*

[4]*Beijing Academy of Quantum Information Sciences, Beijing 100193, China*

[5]*Department of Physics, Southern University of Science and Technology, Shenzhen 518055, China*

The interplay between quantum criticality and Fermi surface reconstruction is central to elucidating the phase diagram of high-temperature cuprate superconductors. While studies on electron-doped T′-structure cuprates suggest an antiferromagnetic origin of this reconstruction, quantitative consensus has been hindered by apical oxygen instabilities and uncontrolled oxygen vacancies. Here, we overcome these limitations by utilizing ozone-assisted molecular beam epitaxy to synthesize high-quality, oxygen-stoichiometric thin films of infinite-layer cuprate $Sr_{1-x}Nd_xCuO_2$ across its entire superconducting dome. Hall transport measurements reveal a sharp carrier-type transition signaling a Fermi surface reconstruction at a critical doping $x_c \approx 0.155$. We show that a spin-density-wave tight-binding model quantitatively reproduces the transport evolution, supporting an antiferromagnetic origin of this quantum phase transition. Furthermore, upon suppressing superconductivity with magnetic fields, the normal-state resistivity exhibits a pristine strange metal behavior that persists down to 2 K in the vicinity of $x_c$. Our findings establish an intrinsic, universal antiferromagnetic quantum criticality in electron-doped cuprates, positioning the structurally simplest infinite-layer cuprates as a clean benchmark platform for theories of unconventional superconductivity.

* To whom correspondence should be addressed. Email: clsong07@mail.tsinghua.edu.cn, zhongyong@quantumsc.cn, xucunma@mail.tsinghua.edu.cn

The doping-dependent ground state of high-temperature (high-$T_c$) cuprate superconductors remains a preeminent challenge in condensed matter physics. At the core of this enigma is the transformation from a parent Mott insulator to a high-$T_c$ superconductor and, ultimately, to a conventional Fermi liquid as the hole doping $p$ (or electron doping $x$) increases [1,2]. The quantum criticality scenario provides a compelling framework for understanding this evolution, claiming that a quantum phase transition at a critical doping $p_c$ (or $x_c$) underlies the anomalous physical properties throughout the phase diagram [3-7]. Over the past two decades, extensive experimental efforts sought the hallmarks of this criticality [8,9]. In hole-doped cuprates such as $YBa_2Cu_3O_{7-\delta}$ and Nd-$La_{2-x}Sr_xCuO_4$, Hall measurements revealed a Fermi surface reconstruction (FSR) across the putative quantum critical point $p_c$ [10,11]. Above $p_c$, the Fermi surface is characterized by a large cylinder whose volume obeys the Luttinger theorem, yielding a carrier density $n = 1 + p$. Below $p_c$, the Fermi surface is folded by the ($\pi$, $\pi$) coherent backscattering, translating into small hole pockets with carrier density $n = p$. In addition, scanning tunneling spectroscopy on $Bi_2Sr_2CaCu_2O_{8+\delta}$ has uncovered that the pseudogap endpoint $p^*$ coincides with the FSR at $p_c$ [12], consistent with the $p$ to $1 + p$ transition observed in $YBa_2Cu_3O_{7-\delta}$ and Nd-$La_{2-x}Sr_xCuO_4$.

Despite these advances, the quantum criticality framework remains a subject of intense debate in hole-doped cuprates. First, the presence of intertwined charge order in the underdoped regime disrupts the $n = p$ relation [6,9,10], complicating the interpretation of FSR as either a primary signature of quantum criticality or a secondary consequence of other electronic orders. Indeed, the detection of small electron-like pockets in the charge order state of $YBa_2Cu_3O_{7-\delta}$ suggests a FSR induced by spatial symmetry-broken order [13]. Second, the universality of the transition from $n = 1 + p$ to $n = p$ in the overdoped regime is contested across different cuprate families [14-16]. Specifically, this transition is not yet definitely established in Bi-based superconductors, leaving a critical gap between transport measurements and spectroscopic observations. Finally, the elusive nature of the pseudogap itself further deepens the challenge of disentangling its interplay with quantum critical phenomena in hole-doped cuprates.

In contrast, electron-doped cuprates provide a cleaner platform for investigating the relationship between quantum criticality and FSR from both experimental and theoretical perspectives [17-19]. Experimentally, the relatively low upper critical field $H_{c2}$ enables direct access to the low-temperature normal states. For instance, angle-resolved photoemission spectroscopy (ARPES), Hall transport, and quantum oscillation measurements in T′-structure $Nd_{2-x}Ce_xCuO_4$ (NCCO) and $Pr_{2-x}Ce_xCuO_4$ (PCCO) have shown that FSR occurs at a critical doping $x_c$, where short-range AFM order vanishes and the carrier density jumps from $n = -x$ to $n = 1 - x$ [20-22]. Theoretically, the suppressed charge order and pseudogap in T′-structure cuprates significantly simplify Hamiltonian modeling by dropping these complex many-body interactions that should be considered in hole-doped cuprates [23,24]. Nevertheless, a quantitative discrepancy persists between the carrier densities extracted from Hall coefficients and those predicted by the standard FSR model [19,25]. This inconsistency is often attributed to the influence of oxygen vacancies ($O_V$), whose concentration is notoriously difficult to control and quantify in T′-structure compounds [25]. Furthermore, given the structural diversity within the cuprate family, whether this AFM-driven FSR picture is a universal feature of electron-doped systems remains an open question. Therefore, the investigation of oxygen-stoichiometric electron-doped cuprates beyond the T′-structure is urgently required to test the correspondence between macroscopic transport signatures and the microscopic mechanism of quantum criticality.

Infinite-layer (IL) $SrCuO_2$ has the simplest structural motif among cuprates, consisting of alternating $CuO_2$ planes and alkaline-earth layers (see inset of Fig. 1(a)). Notably, recent STM measurements on IL cuprates reveal the absence of charge order and pseudogap above $T_c$ [26-28], making them an ideal platform to explore the role of AFM correlations in the formation of FSR and exotic phenomena associated with quantum criticality. While electron doping via $La^{3+}$ substitution prompts superconductivity with $T_c$ exceeding 40 K [29,30], the metastability of the IL phase renders it technically challenging to synthesize $Sr_{1-x}La_xCuO_2$ (SLCO) in the overdoped regime [31,32]. With increasing $La^{3+}$ content ($x > 0.14$), interstitial oxygen progressively occupies the apical sites ($O_A$) and induces a structural transition to the superoxygenated long-$c$ phase [33]. This structural instability has hindered systematic investigations of the quantum critical phenomena in overdoped IL cuprates. To overcome this solubility limit, in this Letter, we report the synthesis of high-quality $Sr_{1-x}Nd_xCuO_2$ (SNCO) thin films on $KTaO_3(001)$ substrates using a home-built ozone-assisted molecular beam epitaxy (MBE) system. By employing the smaller lanthanide ion $Nd^{3+}$ together with a tensile strain from substrate, we successfully extend the electron-doping range to encompass the entire superconducting dome in an IL cuprate system. Through precise optimization of post-growth reduction annealing to achieve strict oxygen stoichiometry for the first time, we comprehensively investigate the doping-dependent evolution of the ground state of IL cuprates. Hall measurements reveal an AFM-driven FSR at $x_c \approx 0.155$, marking a quantum critical point that exhibits characteristic strange metal behavior.

Figures 1(a) and (b) display the $2\theta$-$\omega$ X-ray diffraction patterns and the in-plane resistivity $\rho_{xx}$ for SNCO films across a wide range of Nd doping $x$. The high quality of the epitaxial SNCO films is evidenced by the presence of clear Laue fringes and the absence of impurity phases (see Fig. S1 for details in Supplemental Material [34]). Figure 1(c) shows the $\rho_{xx}(T)$ curves for a $x = 14\%$ SNCO film under out-of-plane magnetic fields. The systematic transport results of a representative sample $A_1$ are exhibited in Figs. 1(c-e), Fig. S2 and Fig. S3. The two-dimensional nature of the superconductivity is confirmed by a cusp-like angle-dependent upper critical field $\mu_0 H_{c2}$ [Fig. S2]. The isotherm $I$-$V$ curve yields a Berezinskii-Kosterlitz-Thouless (BKT) transition temperature of $T_{\mathrm{BKT}} = 20.4$ K, determined by the $V \propto I^3$ power-law criterion [Figs. 1(d) and (e)] [35,36]. This value is also consistent with the $T_{\mathrm{BKT}}$ deduced from the Halperin-Nelson relation [37] [Fig. S3]. Compared with previous reports, the sample quality in this work is significantly improved in several key aspects. First, for $x > 0.144$, the films exhibit well-defined metallic behavior ($\mathrm{d}\rho_{xx}/\mathrm{dT} > 0$) down to 2 K under magnetic fields [Figs. S4 and S5], with no low-temperature resistivity upturns commonly observed in earlier SNCO polycrystals or thin films [26,28-31,38]. Second, the $c$-axis lattice constant decreases linearly with $x$ from the parent compound up to the heavily overdoped regime ($x = 0.203$) [Fig. 1(f)]. This behavior differs from the nonmonotonic variation of $c$ arising from doping-dependent apical oxygen ($O_A$) incorporation [28,39], signifying negligible $O_A$ concentration in our tensile-strained SNCO films. Remarkably, the residual resistivity ratio (RRR) increases rapidly with $x$ and saturates at ~ 2.3 in the overdoped regime. This rules out degraded sample quality as the cause of superconductivity suppression upon overdoping. Third, the absence of oxygen vacancy ($O_V$) is naturally satisfied in the IL structure since the charge reservoir layer has no oxygen atoms. Taken together, these high-quality, oxygen-stoichiometric thin films provide a pristine platform for studying quantum criticality in IL cuprates.

According to the standard FSR model [40], the Fermi surface topology evolves through three distinct regimes as a function of $x$, labelled as Phase I, II and III in Fig. 2(a). Figure 2(b) displays the representative Hall resistivity ($\rho_{xy}$) measurements at various temperatures for each phase. In the normal state, $\rho_{xy}$ varies linearly with the magnetic field $\mu_0 H$ up to 9 T. The Hall coefficient, $R_H$ =

$\rho_{xy}(\mu_0 H)/\mu_0 H$, changes from negative in Phase I to positive in Phase III [Fig. 2(b)], indicating a transition of the dominant carriers from electrons to holes. In Phase II, the $R_H$ exhibits two sign reversals upon cooling, evolving from positive to negative and back to positive, in alignment with the $R_H$ reported in $Sr_{0.88}La_{0.12}CuO_2$ with minimum $O_A$ [41]. This anomalous behavior indicates the coexistence of electron and hole pockets. The observation of doping-dependent carrier-type evolution provides critical evidence for the existence of FSR in IL cuprates.

To precisely locate the FSR transition point $x_c$, Figure 3(a) presents the temperature dependence of $R_H$ across a wide doping range from $x = 0.084$ to $x = 0.203$. In both Phases I and III, the absolute magnitude of $R_H$ decreases monotonically with increasing temperature, a behavior conventionally associated with thermally activated carriers. In contrast, $R_H$ within Phase II exhibits a pronounced temperature dependence in both magnitude and sign, indicative of a multi-pocket electronic structure. To obtain the ground-state evolution, the doping dependence of $R_H$ at 10 K is extracted from $R_{xy}(\mu_0 H)$ in Fig. S6 and plotted in Fig. 3(b). In Phase III, the positive $R_H$ follows the expectation for a conventional large hole-like Fermi surface with carrier density $n = 1 - x$ [Fig. 2(a), right panel]. Conversely, in Phase I, $R_H$ becomes negative and approaches the single-electron-pocket limit ($n = -x$) [Fig. 2(a), left panel]. These behaviors are marked by the solid curves corresponding to $R_H = 1/[(1 - x)eV_B]$ and $R_H = -1/(xeV_B)$, where $V_B$ denotes the volume of the first Brillouin zone.

The critical point $x_c \approx 0.155$, at which $R_H$ departs from the large hole Fermi surface curve, signals the onset of the FSR and the boundary of Phases II and III. At $x_c$, the large hole cylinder at the Brillouin zone corner undergoes a topological reconstruction: an electron pocket emerges around the antinodal points, accompanied by folded hole pockets along the nodal direction [Fig. 2(a), middle panel]. Such behavior is fundamentally incompatible with a rigid-band shift picture, wherein electron doping would merely shrink the hole pocket monotonically. Considering that $x_c$ coincides with the ground-state endpoint of the AFM phase as discussed below, it provides compelling evidence for an FSR driven by (π, π) antiferromagnetic correlations. Given the absence of competing charge order and pseudogap phases in IL SNCO films, these results lead us to conclude an antiferromagnetic quantum critical point (QCP) at $x_c$ – constituting the primary finding of this work.

To gain further insight into the observed antiferromagnetic QCP, we model the electronic structure using a tight-binding dispersion on a square lattice,

$$\epsilon(k_x, k_y) = -2t_1(\cos k_x a + \cos k_y a) + 4t_2 \cos k_x a \cos k_y a - 2t_3(\cos 2k_x a + \cos 2k_y a),$$

combined with a doping-dependent spin-density-wave (SDW) backscattering amplitude $\Delta(x) = \Delta(0)\sqrt{1-\frac{x}{x_c}}$ with wavevector $\boldsymbol{Q} = (\pi, \pi)$, where $t_1$, $t_2$, and $t_3$ represent the nearest-, next-nearest-, and third-nearest-neighbor hopping parameters, respectively [40] (see Supplementary Materials for details). As shown by the pink dashed line in Fig. 3(b), the calculated $R_H(x)$ quantitatively reproduces the experimental data, with the fitting parameters listed in Table 1. The closure of the SDW gap at $x_c \approx 0.155$ aligns with the values reported for T′-structure cuprates [21,22,25,41-44]. The extracted SDW gap $\Delta$ (0) ≈ 0.623 eV is comparable to previous results deduced from Hall coefficient and ARPES studies for T′-structure cuprates [20,40,45]. Furthermore, we identify the boundary separating Phases I and II at $x_h \approx 0.112$, at which the $E_F$ reaches the top of the reconstructed hole-pocket centered at (π/2, π/2).

Having established the presence of antiferromagnetic QCP-driven FSR in IL cuprates, we next examine the temperature dependence of the longitudinal resistivity $\rho_{xx}$ in the normal state. Linear-in-temperature ($\rho_{xx} \propto T$) and linear-in-field ($\rho_{xx} \propto H$) behaviors are hallmarks of the strange-metal state and are widely regarded as signatures of quantum criticality in strongly correlated systems [19,46-52]. As shown in Figs. 4(a) and S7, the normal-state resistivity of SNCO films near optimal doping exhibits linear-in-temperature dependence from ~ 40 K down to $T_{c,\,onset}$. Upon suppressing superconductivity with an out-of-plane magnetic field, this resistivity linearity extended to the lowest measured temperatures, when a linear-in-field ($\mu_0 H$) magnetoresistance response is considered. To evaluate this behavior, we use the combined temperature and field dependence [52-54],

$$\rho_{xx}(T,H) - \rho_{xx}(0,0) = \sqrt{(\alpha k_B T)^2 + (\beta \mu_B \mu_0 H)^2},$$

the experimental data can be well fitted, as indicated by orange dashed lines. The ratio $\beta/\alpha$ is of order unity and shows weak dependence on $T_c$ near $x_c$, consistent with the universal QCP behavior observed in cuprates and iron-based superconductors [55]. As the system moves away from $x_c$ toward the Fermi-liquid regime, the quadrature scaling breaks down: the ratio $\beta/\alpha$ deviates significantly from unity, accompanied by a suppression of the $T$-linear coefficient $\alpha$.

To further elaborate on the quantum criticality, we fit the temperature-dependent resistivity up to 100 K using a phenomenological form [19,49,56],

$$\rho_{xx} = a_0 + a_1 T + a_2 T^2,$$

where $a_0$, $a_1$ and $a_2$ denote temperature-independent coefficients. This polynomial form captures the combined contributions from strange-metal ($T$-linear) and Fermi-liquid ($T^2$) components and well describes all experimental data near and above the optimal doping [Fig. S8]. In contrast, this phenomenological form breaks down in underdoped samples due to the low-temperature resistivity upturn. Figure 4(b) shows the extracted value of $a_1/a_2$. In the vicinity of QCP, the ratio $a_1/a_2$ is much larger than unity, indicating dominant strange-metal behavior. Away from the QCP, $a_1/a_2$ decreases rapidly, signaling a crossover to Fermi-liquid transport. In the heavily overdoped regime, $a_1/a_2 \ll 1$, consistent with conventional quasiparticle scattering.

Based on the present study and previous reports, we summarize the electronic phase diagram of electron-doped cuprates in Fig. 5. Both $T_{c,\,onset}$ and $T_{c,\,zero}$ exhibit a dome-shaped dependence on the Nd content $x$, having an optimal doping at $x_{opt} \approx 0.143$ and a maximum $T_{c,\,onset} \approx 28$ K [Fig. S5, [34]], in excellent agreement with previous STM observations [26]. Two critical doping levels, $x_h$ and $x_c$, separate the phase diagram into three distinct regions, corresponding to the three phases in Fig. 2(a). The FSR transition point $x_c \approx 0.155$ has been obtained in IL cuprates by both experimental measurements and tight-binding simulations. Notably, this value aligns well with the quantum critical point range ($x_c \approx 0.145$-$0.170$) typically reported for T′-structure cuprates [21,22,42,45,57,58]. Intriguingly, the lower critical doping $x_h$ coincides with the onset of the superconducting dome, highlighting the key role of hole pockets for the emergence of superconductivity in electron-doped cuprates. This is consistent with previous observations in T′-structure systems, where quantities such as the Hall angle, superfluid density, and Fermi-surface spectral weight correlate $T_c$ with the emergence of the hole pocket [59-61]. The recurrence of this behavior across different electron-doped cuprates points to a potentially universal connection between hole pockets and superconductivity in cuprates.

Figure 5 displays the temperature-dependent phase boundaries for both long- and short-range AFM orders as well. In T′-structure compounds such as $(Pr, La)_{2-x}Ce_xCuO_{4\pm\delta}$ (PLCCO) and NCCO, extensive neutron scattering and nuclear magnetic resonance (NMR) studies have shown that static long-range AFM order terminates slightly below the critical point $x_c$ [62,63]. Although neutron scattering data are currently lacking for IL cuprates, similar magnetic behavior can be inferred from resistivity anomalies in transport measurements. In the underdoped regime, a distinct resistivity hump is often attributed to a static AFM transition [16,28,64,65]. This hump progressively weakens with increasing doping and vanishes around $x_c \approx 0.120$ [Fig. S8], slightly below $x_c$, reminiscent of the trend observed in T′-structure systems [57,61]. While this long-range AFM order collapses prior to $x_c$ at zero field, muon-spin relaxation ($\mu$SR) measurements on PLCCO and NCCO demonstrate that a short-range spin-freezing state or short-range AFM fluctuations persist up to the vicinity of $x_c$ [61,66]. Additionally, angle-dependent magnetoresistance measurements on LCCO and PCCO reveal that short-range AFM order survives deep within the superconducting dome and can be extrapolated precisely to $x_c$ at zero temperature [67,68]. Notably, recent NMR investigations on Eu-$La_{2-x}Sr_xCuO_4$ have shown that a strong magnetic field is capable of stabilizing short-range AFM fluctuations into static long-range order up to the putative quantum critical point $p_c$ [69]. Therefore, the magnetic fields utilized in our Hall transport measurements can effectively extend the phase boundary of the static AFM order, shifting its termination point to the FSR critical point $x_c$. Here, we emphasize that while the FSR captured in zero-field ARPES studies is typically attributed to short-range AFM order [58], its precise correspondence to high-field Hall transport and magnetoresistance measurements needs further investigation.

In summary, we realize high-quality epitaxial infinite-layer $Sr_{1-x}Nd_xCuO_2$ thin films spanning the entire superconducting dome, achieving unprecedented stoichiometry control in electron-doped infinite-layer cuprates. This enables direct access to intrinsic magneto-transport, revealing an AFM-driven Fermi surface reconstruction at a QCP around which a characteristic strange metal behavior emerges. Importantly, a quantitative agreement has been established between the macroscopic Hall transport and the microscopic Hamiltonian model. The universal electronic structure evolution shared with the T′-structure demonstrates a common phase diagram for electron-doped cuprates. Therefore, our work positions the structurally simplest infinite-layer cuprate as a unique platform to elucidate quantum criticality and place stringent constraints on the mechanism of unconventional superconductivity.

**Acknowledgments**

This work is financially supported by the National Key R&D Program of China (2022YFA1403100), the Natural Science Foundation of China (12134008, 12474130, 12141403, 52388201), and the Innovation Program for Quantum Science and Technology (2021ZD0302502).

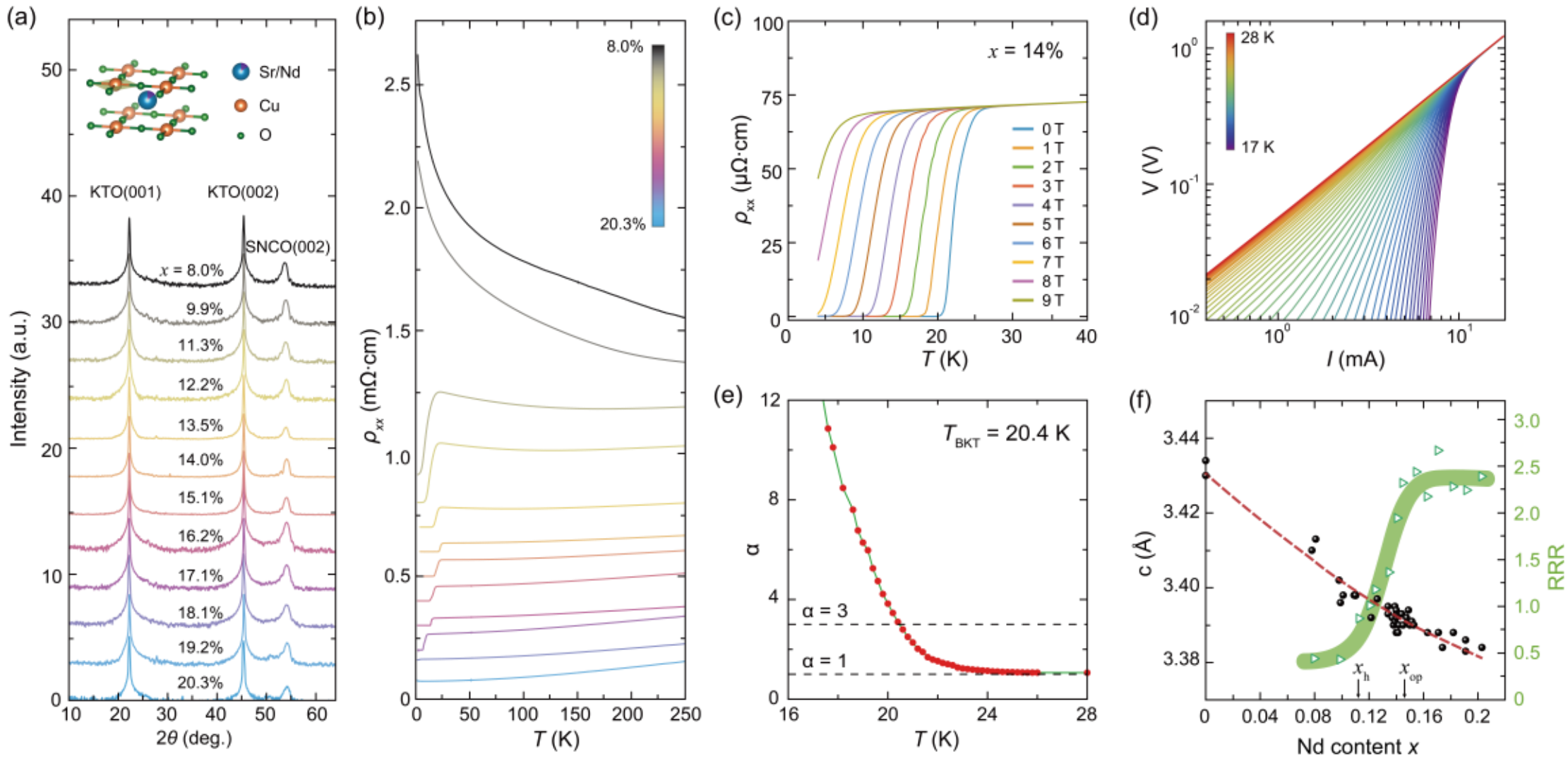

FIG. 1. (a) $2\theta$-$\omega$ X-ray diffraction scans of SNCO thin films with various Nd content $x$. (b) Temperature dependence of the in-plane resistivity $\rho_{xx}$ for various doping levels $x$. The curves are vertically offset for clarity. (c) In-plane resistivity of a slightly underdoped SNCO sample ($x$ = 14%) under out-of-plane ($H//c$) magnetic fields. (d) Isotherm $I$-$V$ curve with temperatures across the superconducting transition region. (e) Temperature dependence of the power-law exponent $\alpha$ derived from $V \propto I^{\alpha}$ scaling relation in (d). (f) Out-of-plane lattice constant $c$ (left axis, black dots) and residual resistivity ratio RRR = $R$(300 K)/$R$(30 K) (right axis, green triangles) as a function of $x$. The red line and the green line serve as a guide to the eye.

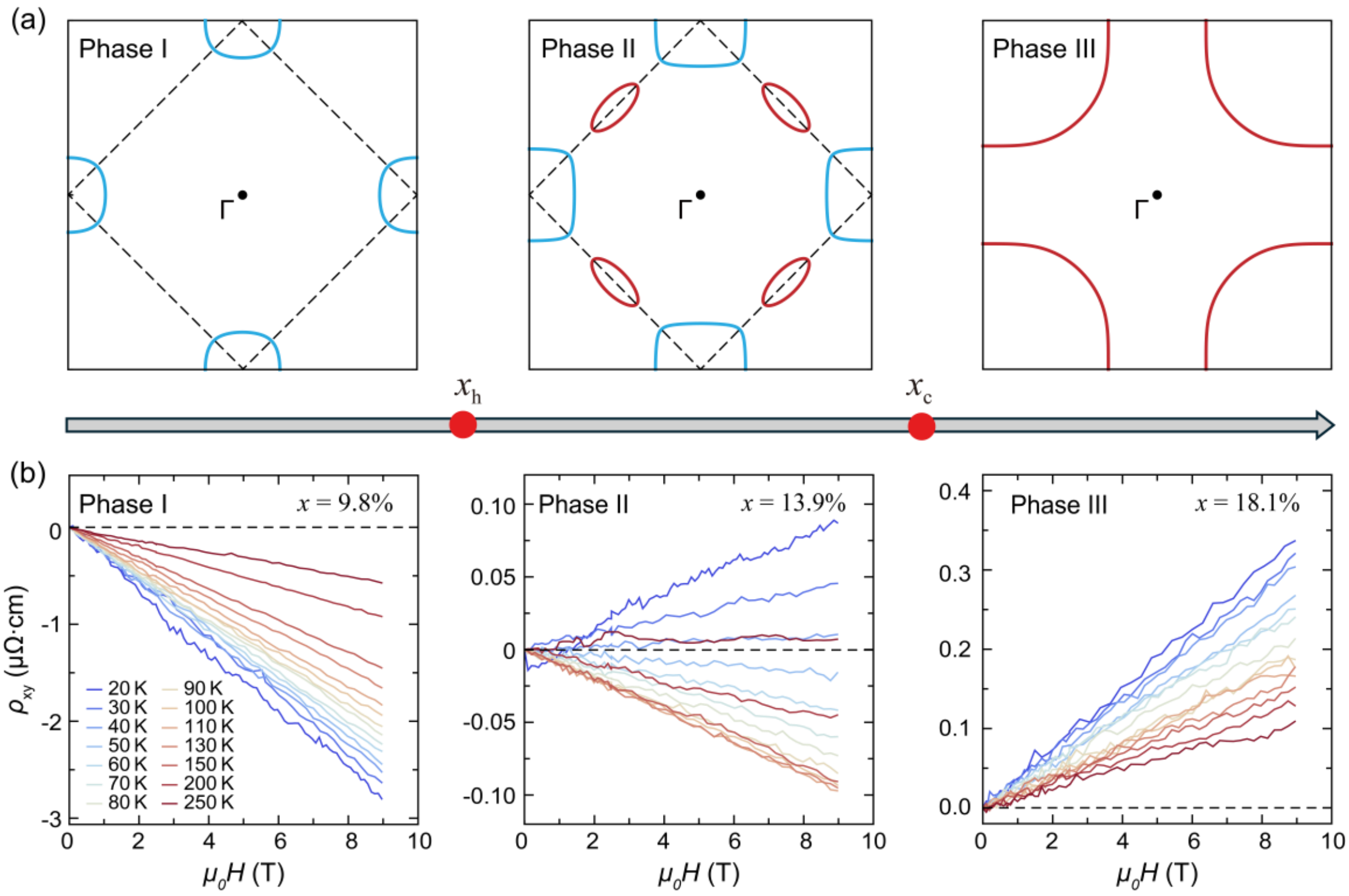

FIG. 2. (a) Schematic evolution of the FS in electron-doped cuprates from the underdoped (Phase I) to optimally doped (Phase II) and overdoped (Phase III) regimes. The reconstructed FS consists of an isolated electron pocket in Phase I, coexisting electron and hole pockets in Phase II, and a large hole-like Fermi surface in Phase III. (b) Magnetic field-dependent Hall resistivity $\rho_{xy}$ at selected temperatures for three representative SNCO films with $x$ = 9.8% (Phase I), $x$ = 13.9% (Phase II), and $x$ = 18.1% (Phase III).

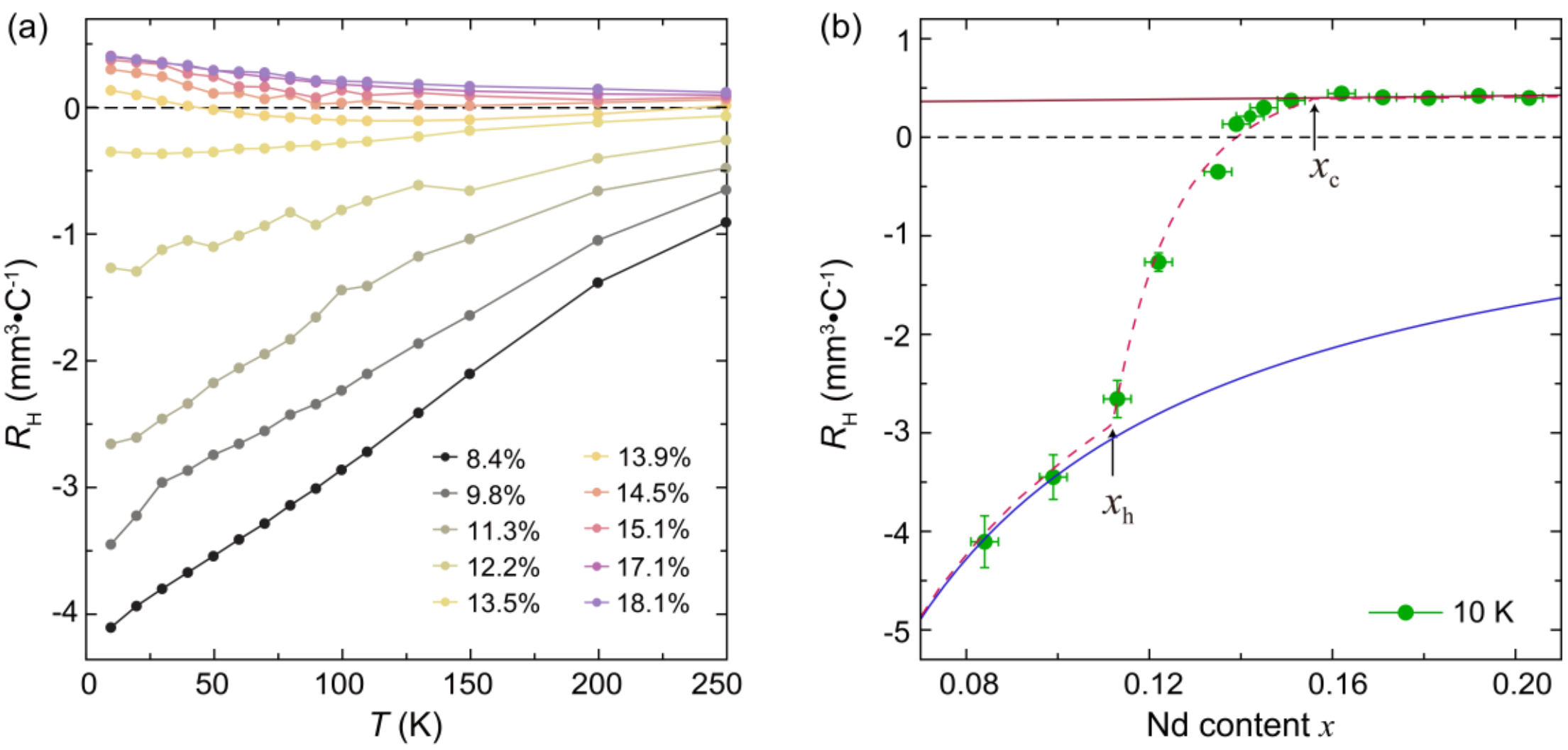

FIG. 3. (a) Temperature dependence of $R_H$ for SNCO films with various Nd content $x$. (b) $R_H$ at 10 K as a function of Nd contents $x$. Horizontal error bars reflect doping uncertainties from flux fluctuations during growth; vertical error bars represent uncertainties in film thickness. The claret red and azure blue curves depict the expected zero-temperature $R_H$ for a large hole-like FS ($R_H \propto 1/(1 - x)$) and electron-pocket contributions ($R_H \propto -1/x$), respectively. The pink dashed curve shows the results of the SDW-based tight-binding model, described in the text. Note that $x_c$ marks the QCP where the gap opens, and $R_H$ deviates from the $1/(1-x)$ curve, while $x_h$ denotes the onset of the reconstructed hole pocket at (π/2, π/2).

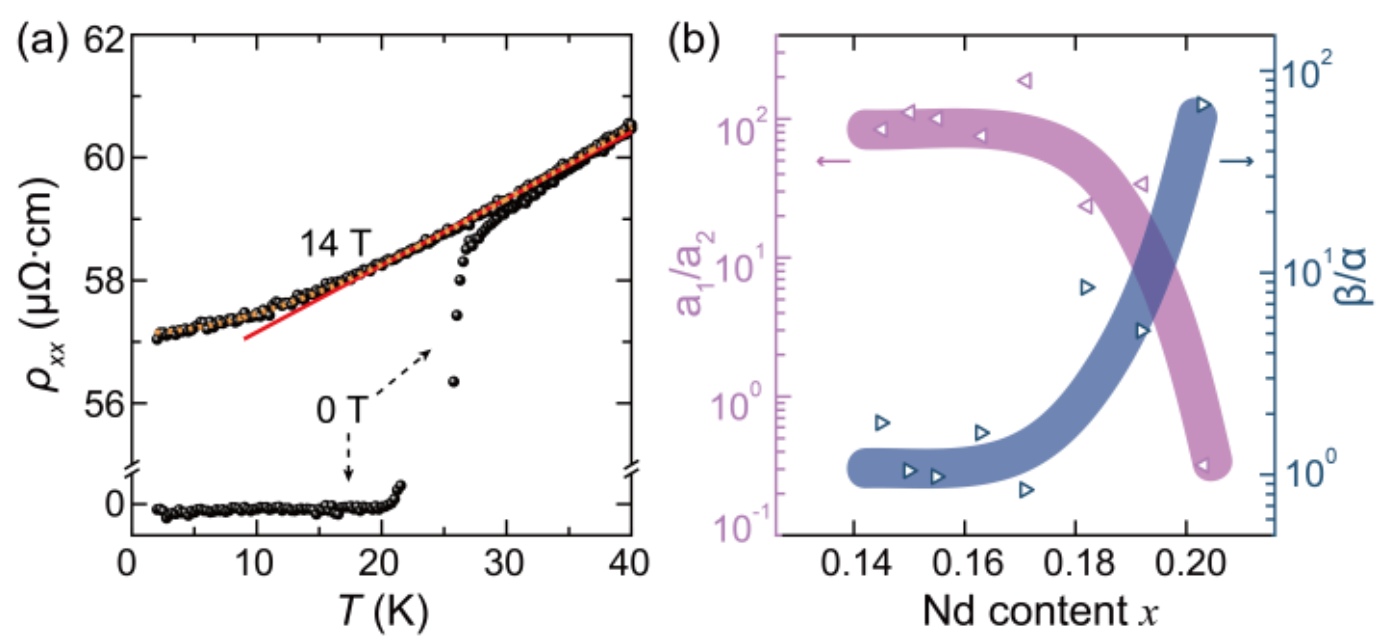

FIG. 4. (a) Temperature dependence of the in-plane resistivity $\rho_{xx}$ for the SNCO film with $x = 0.145$. Upon suppressing superconductivity with magnetic fields, the normal-state resistivity exhibits a robust resistivity linearity down to 2 K, when a linear-in-field ($\mu_0 H$) magnetoresistance response is considered. The orange dashed line is the fit to the scaling relation: $\rho_{xx}(T,H) - \rho_{xx}(0,0) = \sqrt{((\alpha k_B T)^2 + (\beta \mu_B \mu_0 H)^2)}$. The red solid line is the linear fit. (b) Statistical analysis of the fitted parameters. Pink triangles represent the ratio of the linear-to-quadratic coefficients ($a_1/a_2$) obtained from a polynomial fit, $\rho_{xx}(T,0) = a_0 + a_1 T + a_2 T^2$. Blue triangles denote the ratio of the magnetic-to-thermal coefficients ($\beta/\alpha$) extracted from the field-temperature scaling analysis in (a).

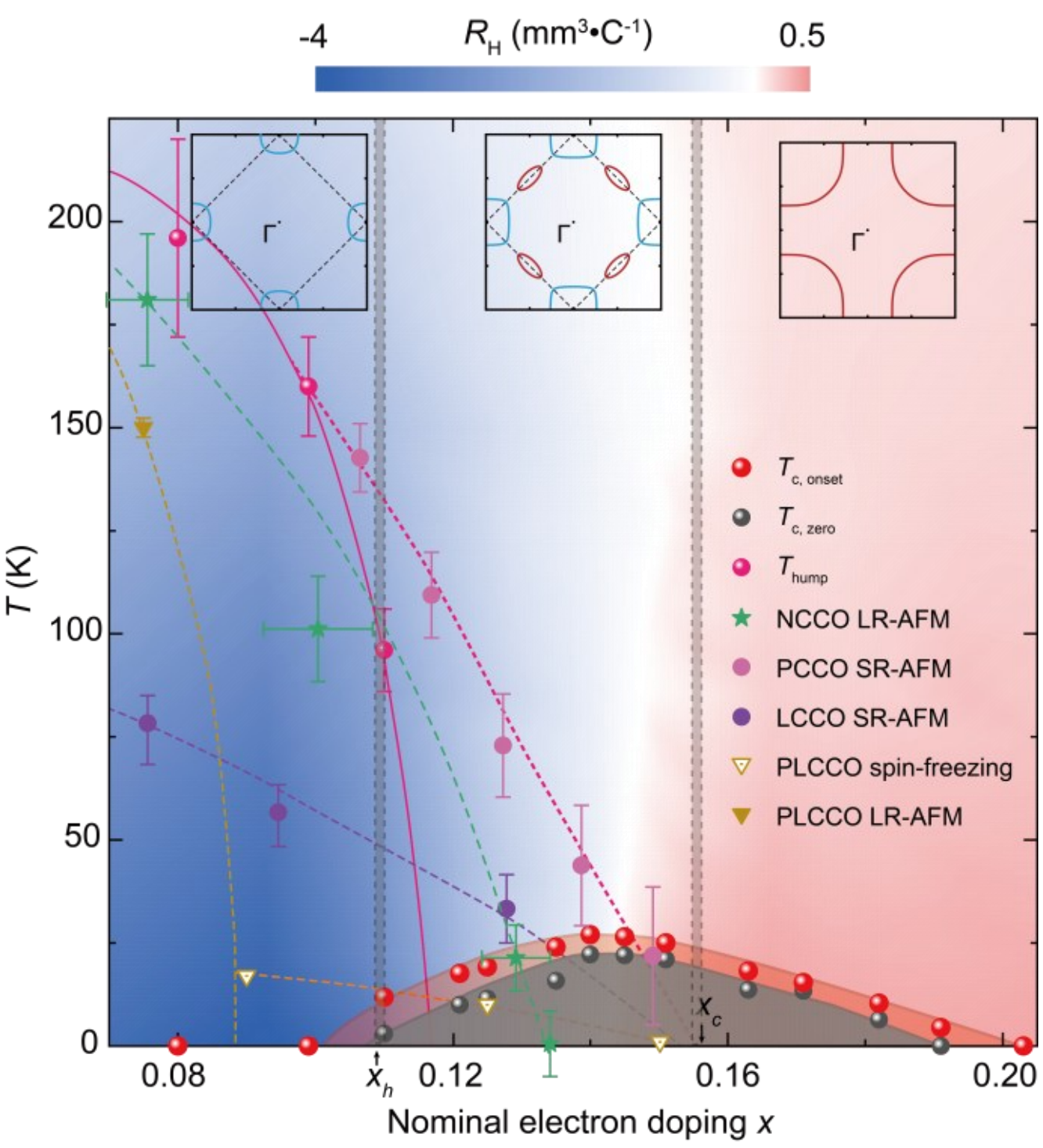

FIG. 5. Electronic phase diagram of SNCO. The superconducting transition temperatures $T_{c,\,onset}$ and $T_{c,\,zero}$ are defined as 90% and 10% of the normal-state resistivity, respectively. For comparison, literature data for related electron-doped cuprates are included: pink and purple solid circles denote the AFM phase boundaries for PCCO [68] and LCCO [67] determined by angle-dependent magnetoresistance; green stars and brown inverted triangles represent AFM phase boundaries extracted from neutron scattering on NCCO [62] and μSR on PLCCO [61], respectively.

Table 1 Fitting parameters for the Hall coefficient

| $t_1$ (eV) | $t_2$ (eV) | $t_3$ (eV) | $x_c$ | $\Delta(0)$ (eV) |
|---|---|---|---|---|
| 0.324(8) | 0.0651(16) | 0.0906(36) | 0.1550(14) | 0.623(20) |

*Supplementary Materials for*

**Antiferromagnetic Quantum Criticality in Infinite-Layer Cuprates $Sr_{1-x}Nd_xCuO_2$**

Bin-Jie Wu[1], Ze-Xian Deng[1], Guang-Ran Wei[1], Xi Zhou[1], Ji-Hai Zhang[1], Da-Zhuang Deng[1], Qun Zhu[1], Yong Zhong[2*], Can-Li Song[1,3*], Xu-Cun Ma[1,3*], Qi-Kun Xue[1,3,4,5]

**Sample growth**

All samples are grown on pretreated insulating $KTaO_3$ (001) substrates using a home-built ozone-assisted molecular beam epitaxy (O-MBE) system integrated with an ozone gas delivery system and reflection high-energy electron diffraction (RHEED) for *in-situ* real-time monitoring. The fluxes of high-purity copper (99.9999%), neodymium (99.9%), and strontium (99.99%) metal sources were precisely calibrated in sequence by using a standard quartz crystal microbalance (QCM, Inficon SQM160H). Prior to growth, the $KTaO_3$ substrates were degassed at 550°C for 10 minutes. The epitaxial films are subsequently grown by co-depositing all metal sources at 565°C under an ozone atmosphere of $4.0 \times 10^{-6}$ Torr. The studied SNCO films s have thicknesses ranging from 7.5 nm to 30 nm, all prepared at a growth rate of approximately 0.44 nm per minute. To overcome the narrow growth window [1], the (Sr+Nd)/Cu flux ratio was controlled within 0.03 to optimize the growth conditions for each doping level. The as-grown films were further post-annealed at 560°C with a base pressure of $3 \times 10^{-9}$ Torr to achieve optimal superconductivity. The excellent ambient stability of SNCO films ensures the intrinsic nature of the XRD and transport measurements [2]. The $2\theta$-$\omega$ scans were conducted using monochromatic copper $K_{\alpha1}$ radiation with a wavelength of 1.5406 Å.

**Transport measurements**

Transport measurements from 2 K to 300 K were carried out using commercial physical property measurement systems (PPMS-9 T and PPMS Dynacool-14 T, Quantum Design). The films are patterned to a standard Hall bar geometry by $Ar^+$ ion etching. For the angle-dependent isothermal magnetoresistance measurements, the sample was mounted on a rotation probe with an angle precision of 0.053°. Unless otherwise specified, an excitation current of 10 μA was used for all measurements.

**Simulation of doping-dependent Hall coefficient**

For the tight-binding model on a two-dimensional square lattice, the electronic dispersion is given by

$\epsilon(k_x, k_y) = -2t_1(\cos k_x a + \cos k_y a) + 4t_2 \cos k_x a \cos k_y a - 2t_3(\cos 2k_x a + \cos 2k_y a)$.

Although such a band structure cannot fully capture the microscopic electronic dispersion in strongly correlated systems, it has been shown that a weak-coupling tight-binding framework incorporating coherent backscattering induced by spin-density wave (SDW) order can effectively reproduce both spectroscopic and transport properties of electron-doped cuprates. In this picture, coherent backscattering with wavevector $\boldsymbol{Q} = (\pi, \pi)$ opens an SDW gap, which is assumed to be doping-dependent: $\Delta(x) = \Delta(0)\sqrt{1 - \frac{x}{x_c}}$, where $x_c$ denotes the QCP associated with the FSR. Diagnosing the SDW Hamiltonian:

$$\begin{pmatrix} \epsilon(\boldsymbol{k}) & \Delta(x) \\ \Delta(x) & \epsilon(\boldsymbol{k}+\boldsymbol{Q}) \end{pmatrix} \vec{P} = E\vec{P},$$

yields two reconstructed bands with dispersions

$$E_{\pm}(\boldsymbol{k}, x) = \frac{1}{2}\left(\epsilon(\boldsymbol{k}) + \epsilon(\boldsymbol{k}+\boldsymbol{Q}) \pm \sqrt{\left(\epsilon(\boldsymbol{k}) - \epsilon(\boldsymbol{k}+\boldsymbol{Q})\right)^2 + 4\Delta(x)^2}\right).$$

The chemical $\mu$ is determined by the electron filling $n = 1 + x$, and the Fermi surface is defined by $E_{\pm} - \mu = 0$. Within the weak-field limit and under the constant scattering-rate approximation, the longitudinal and Hall conductivities are given by [3]

$$\sigma_{xx} = \frac{e^2}{\hbar}\oint \frac{1}{4\pi^2} l(s) ds,$$

$$\sigma_{xy} = \frac{e^2}{\hbar}\frac{B}{\Phi_0}\oint \frac{1}{4\pi} d\vec{l} \times \vec{l} \cdot \hat{\boldsymbol{z}},$$

where $\Phi_0$ is the flux quantum, $s$ denotes the coordinate along the Fermi surface, and $\vec{l} = \overrightarrow{v_F}\tau$ is the mean free path. The fitting parameters are summarized in Table 1. From the fitting parameters, we obtained $x_h \approx 0.112$, corresponding to the emergence of the reconstructed hole pocket centered at (π/2, π/2).

**2D transport nature of SNCO films**

Figure S3 shows the temperature-dependent resistivity under various magnetic fields. The upper critical field $H_{c2}(T)$ is defined as the magnetic field at which the resistivity decreases to 50% of the normal-state value just above $T_c$. The temperature dependence of the upper critical fields for magnetic fields applied perpendicular and parallel to the $CuO_2$ planes exhibits pronounced anisotropy and can be fitted by the 2D Ginzburg-Landau (GL) model [Fig. S2]:

$$\mu_0 H_{c2}^{\perp} = \frac{\Phi_0}{2\pi\xi_{ab}^2}(1-\frac{T}{T_c}),$$

$$\mu_0 H_{c2}^{\parallel} = \frac{\sqrt{12}}{2\pi d_{sc}\xi_{ab}}\sqrt{1-\frac{T}{T_c}},$$

where $\Phi_0 = h/2e$ is the superconducting flux quanta. Specifically, Figure S2(b) presents the extracted $\mu_0 H_{c2}$ as a function of polar angle $\theta$, showing a cusp-like feature near $\theta = 90°$. Consistent with the behavior expected for a 2D superconductivity, the $\mu_0 H_{c2}$ ($\theta$) data are well described by the 2D Tinkham model $\left|\frac{H_{c2}(\theta)\cos\theta}{H_{c2,c}}\right| + \left(\frac{H_{c2}(\theta)\sin\theta}{H_{c2,ab}}\right)^2 = 1$, shown as the red solid line, rather than by the 3D anisotropic Ginzburg-Landau model $\left(\frac{H_{c2}(\theta)\cos\theta}{H_{c2,c}}\right)^2 + \left(\frac{H_{c2}(\theta)\sin\theta}{H_{c2,ab}}\right)^2 = 1$, represented by the blue dashed line. Moreover, the zero-field temperature-dependent resistivity can be fitted using the Halperin-Nelson relation $\rho = \rho_0 \cdot \exp\left(\frac{-b_{HN}}{\sqrt{T-T_{BKT}}}\right)$, where $T_{BKT}$ denotes the BKT transition temperature and naturally corresponds to the $T_{c,\,zero}$ [4]. The consistency between the $T_{BKT}$ extracted from the isothermal *I-V* curves and from the Halperin-Nelson fitting further supports the 2D nature of the observed superconductivity.

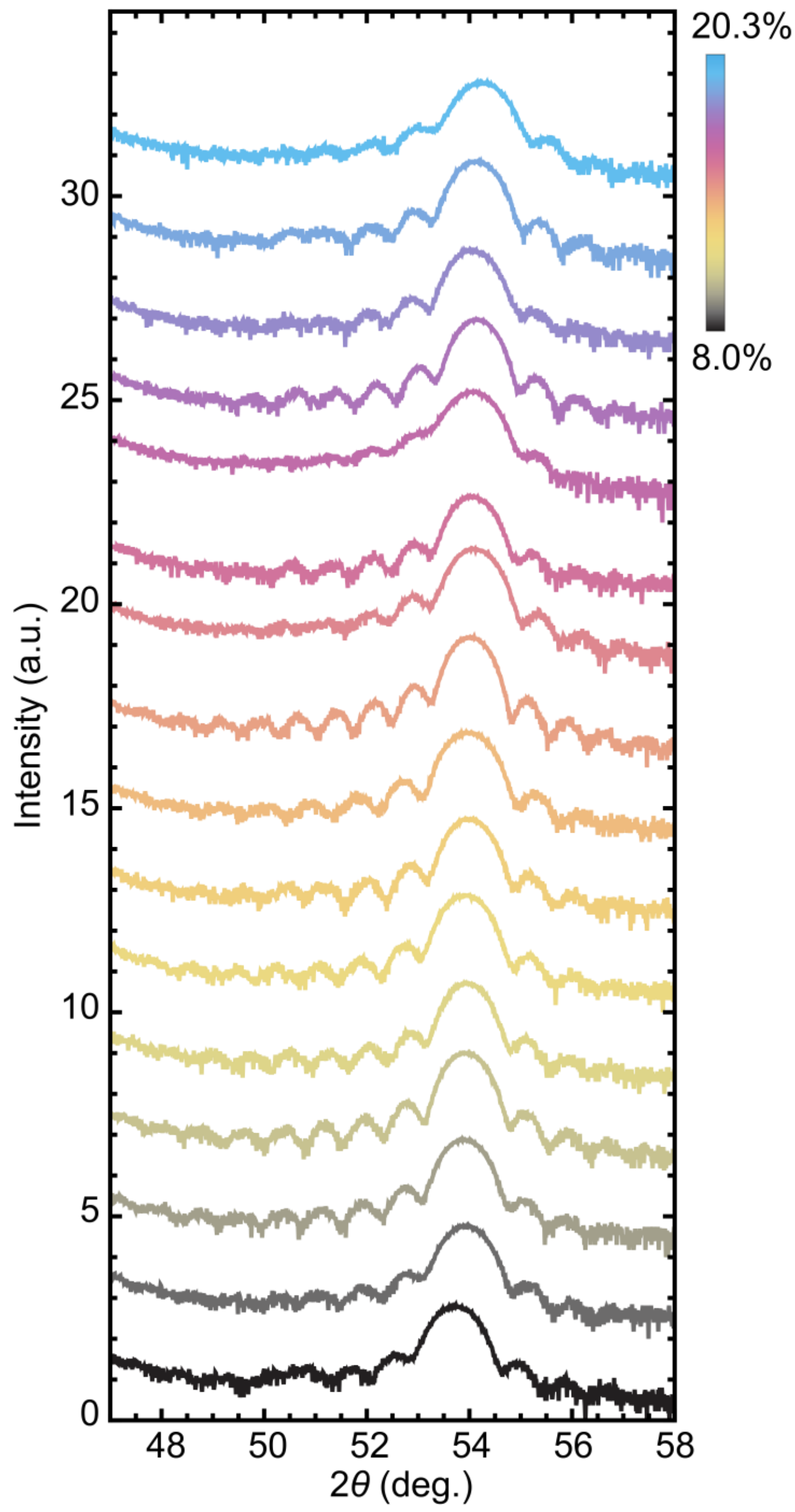

Fig. S1. 2$\theta$-$\omega$ scans of SNCO films with various Nd content $x$, shown in a narrow angular range around the SNCO(002) peaks.

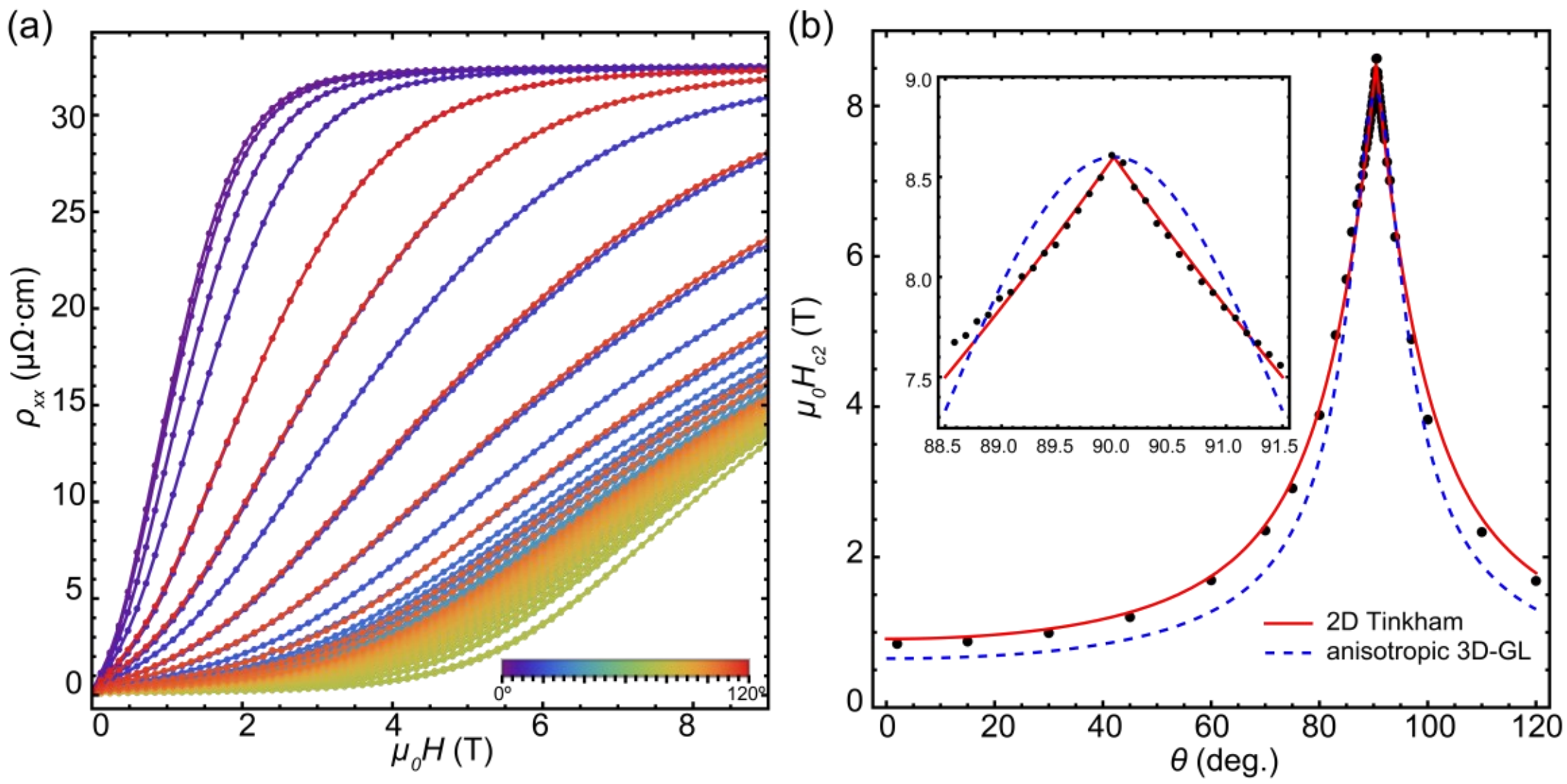

Fig. S2. (a) Polar-angle dependence of the magnetoresistance of sample $A_1$ measured at 20 K. (b) Polar-angle dependence of the upper critical field $H_{c2}$ extracted from (a) using the 40% normal-state resistivity criterion. The red solid and blue dashed curves represent fits to the 2D-Tinkham model and the anisotropic 3D-Ginzburg-Landau model, respectively.

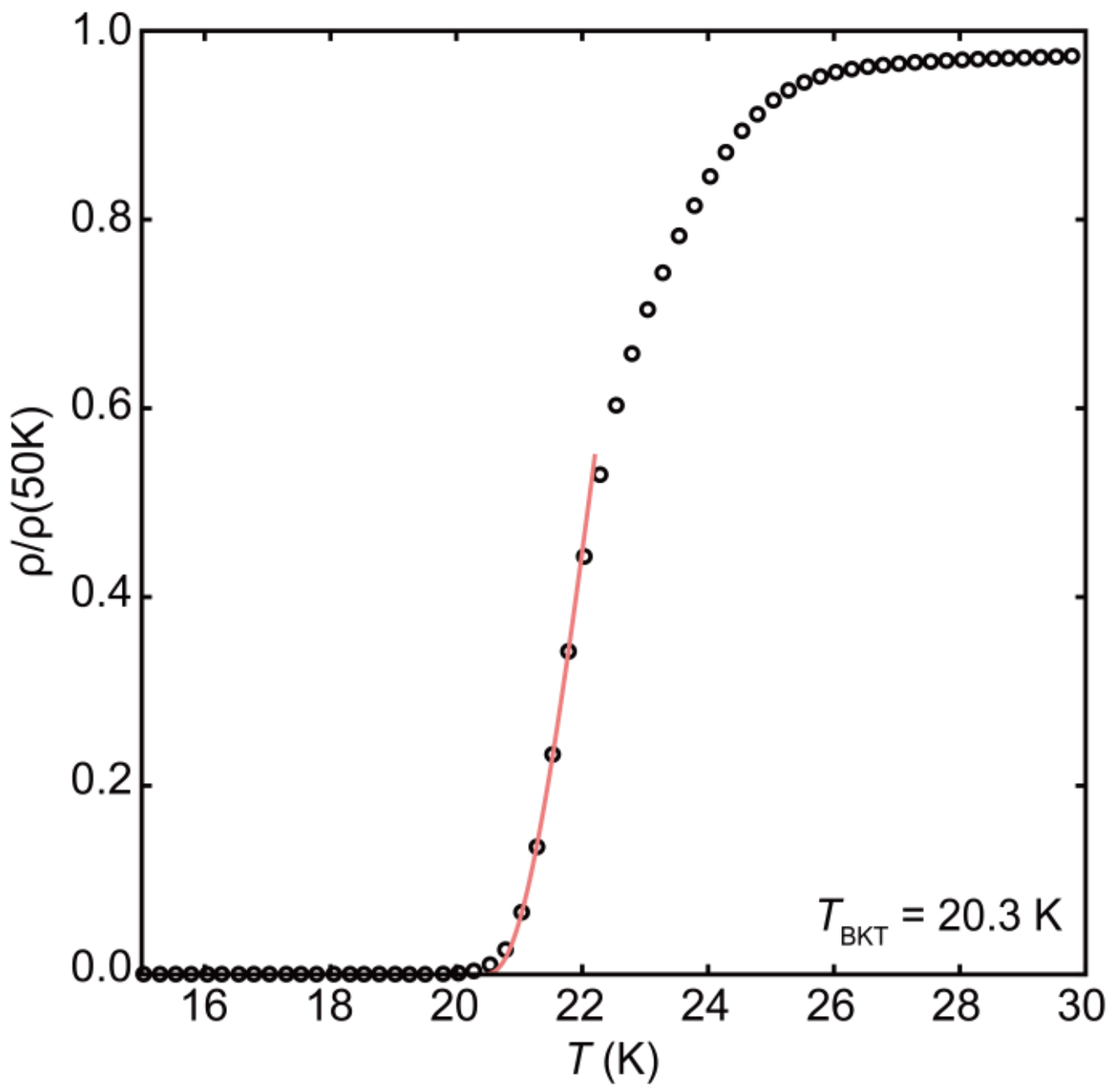

Fig. S3. Halperin-Nelson fit to the resistivity of sample $A_1$.

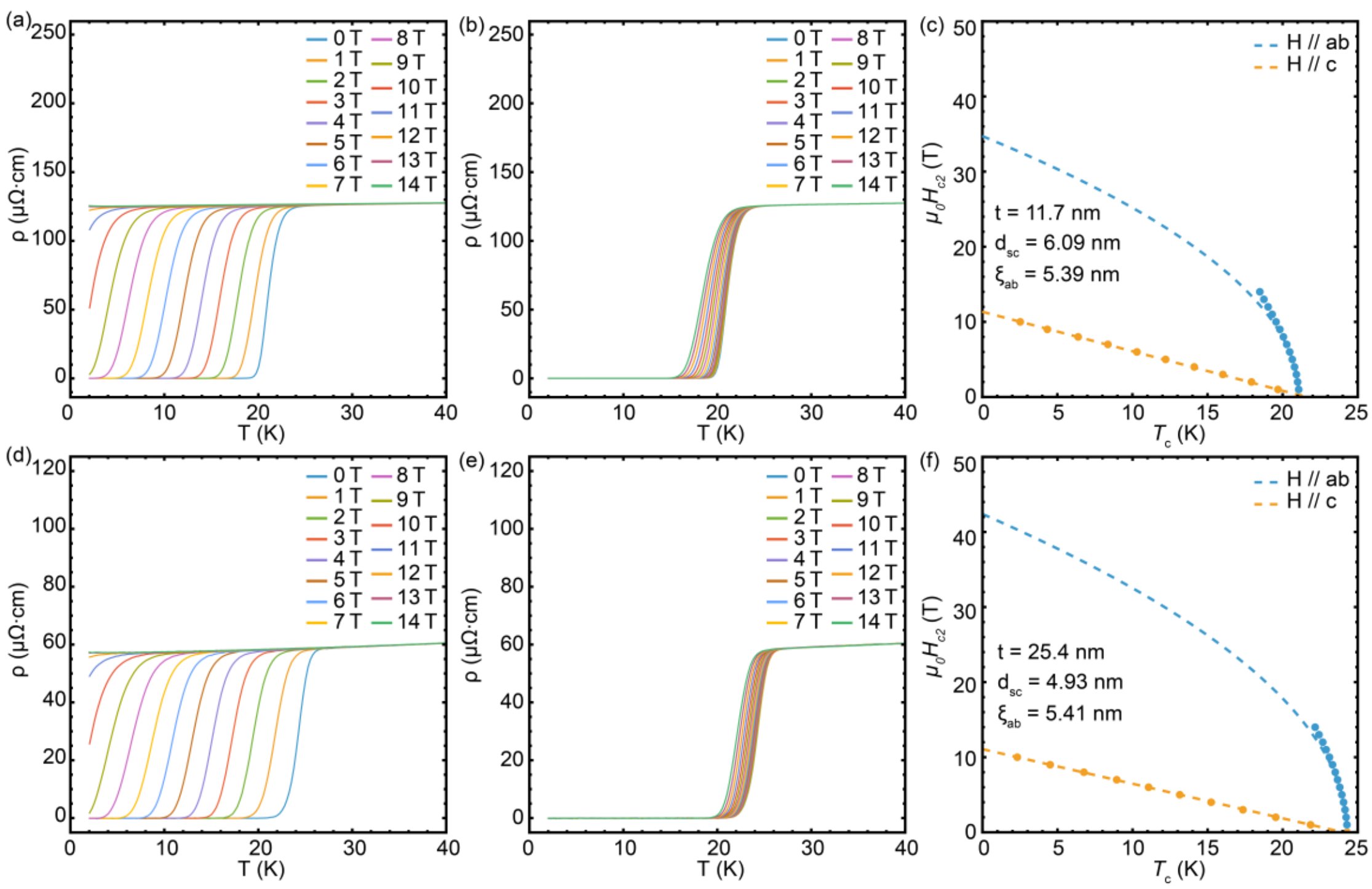

Fig. S4. (a,b) In-plane resistivity of sample $A_2$ ($x$ = 14.0%, 11.7 nm) under (a) out-of-plane ($H//c$) and (b) in-plane ($H//ab$) magnetic fields. (c) Temperature dependence of the upper critical fields $H_{c2}$ extracted from (a) and (b) using the 50% of the normal-state resistivity criterion. (d-f) Same to (a-c), but for another sample $A_3$ (14.4%, 25.4nm).

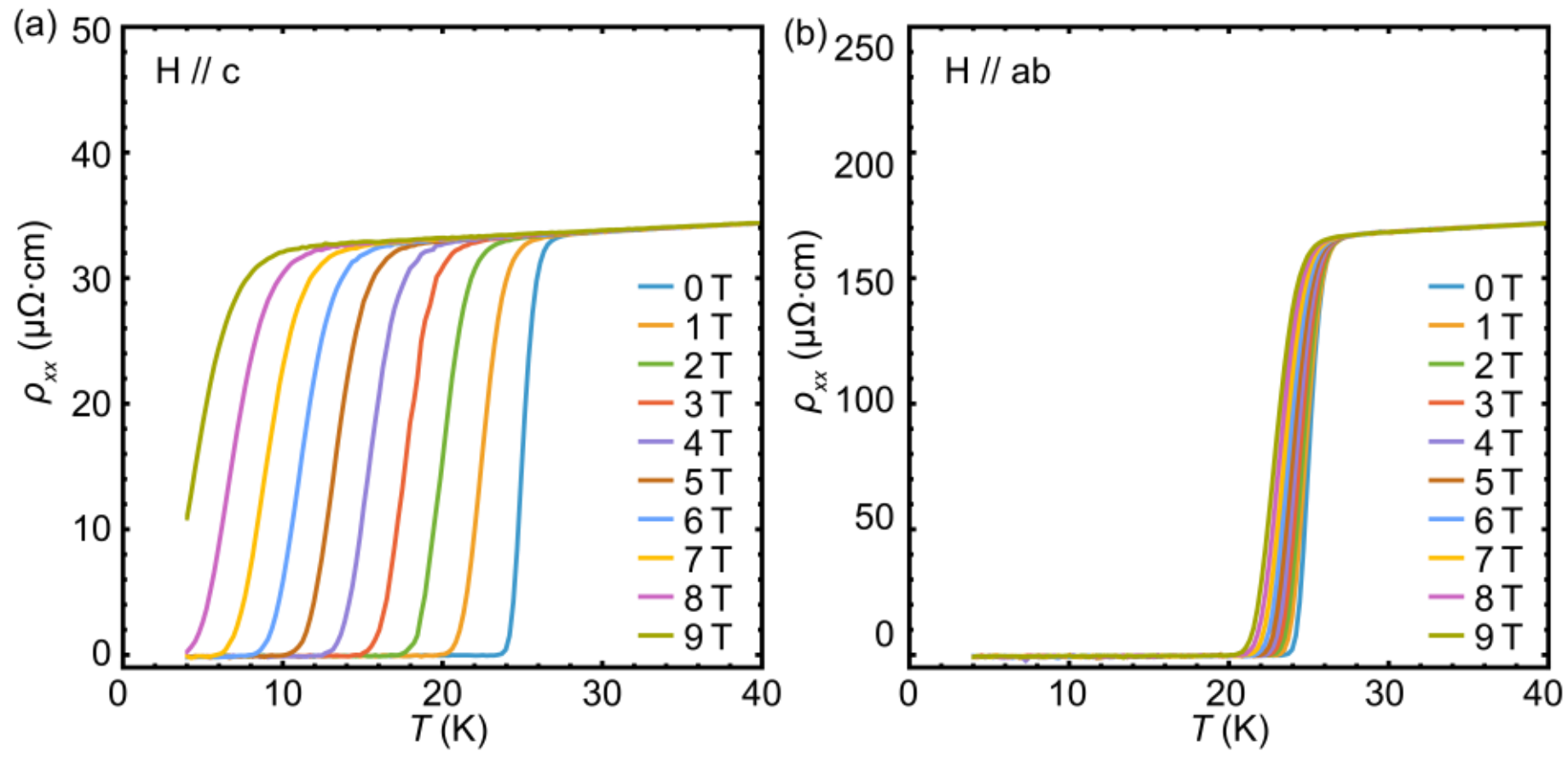

Fig. S5. In-plane resistivity under (a) out-of-plane (*H*//*c*) and (b) in-plane (*H*//*ab*) magnetic fields for the optimally doped SNCO sample $A_4$ (14.4%, 12.1nm).

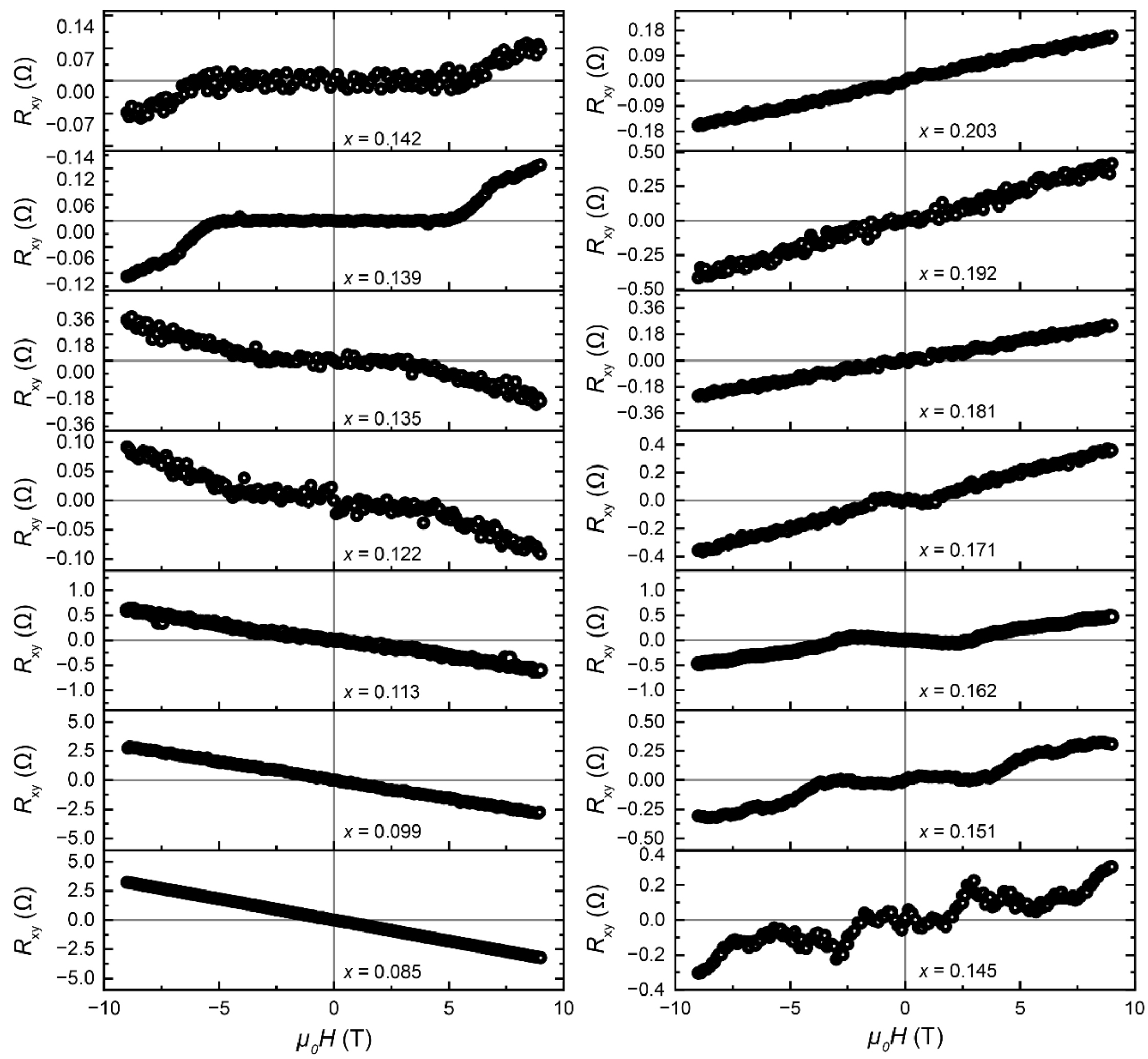

Fig. S6. Symmetrized Hall resistivity measured at 10 K for SNCO samples with various Nd contents $x$. The near-zero Hall resistivity plateau around optimal doping arises from the superconducting state.

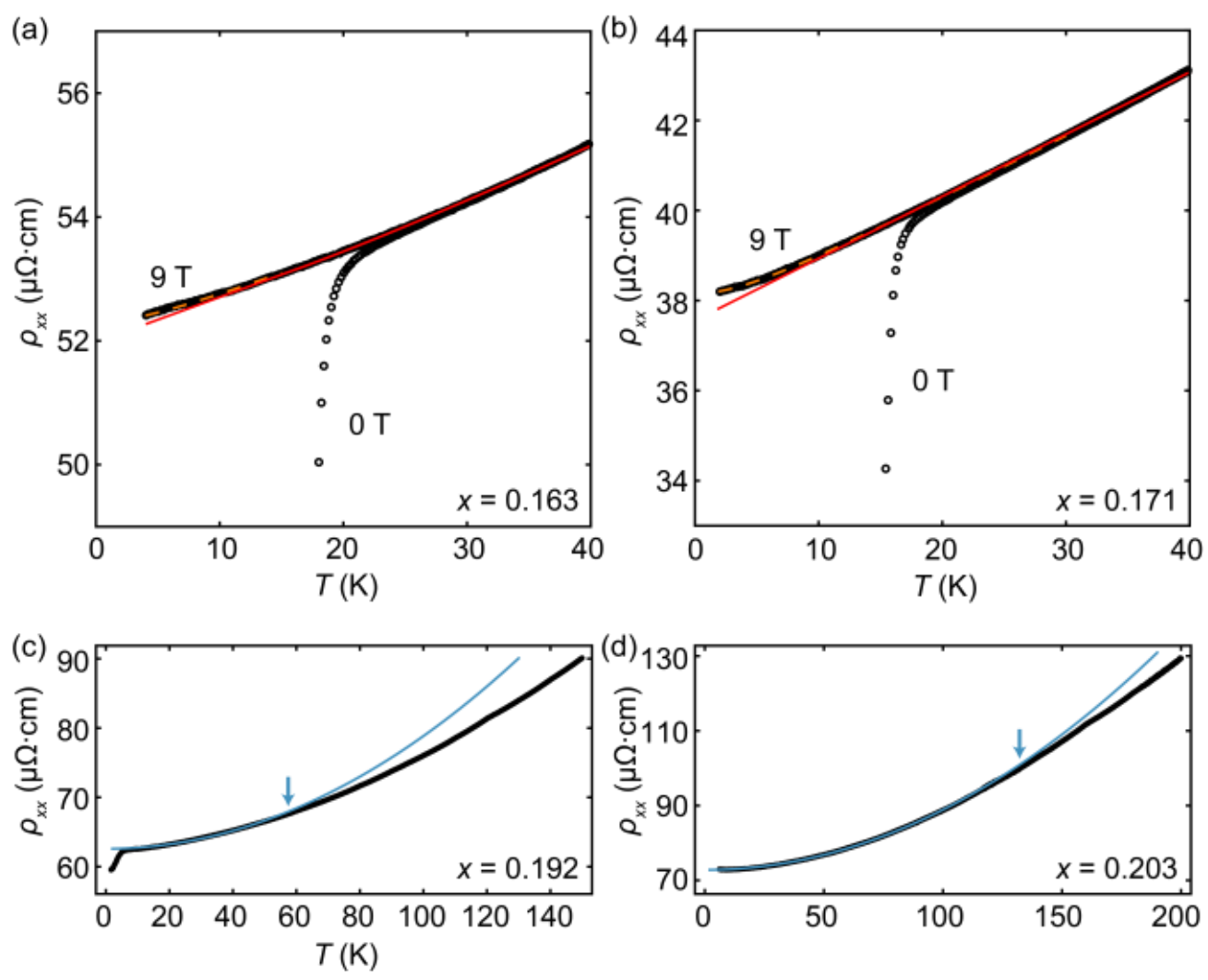

Fig. S7. (a-d) Scaling analyses for SNCO samples with different Nd contents as indicated. The orange curves in (a) and (b) show the best fits of the experimental data to $\rho_{xx}(T,H) - \rho_{xx}(0,0) = \sqrt{(\alpha k_B T)^2 + (\beta \mu_B \mu_0 H)^2}$, while the red solid lines represent linear fits of the data, as in Fig. 4(a). The blue solid curves in (c) and (d) show parabolic fits of the corresponding data.

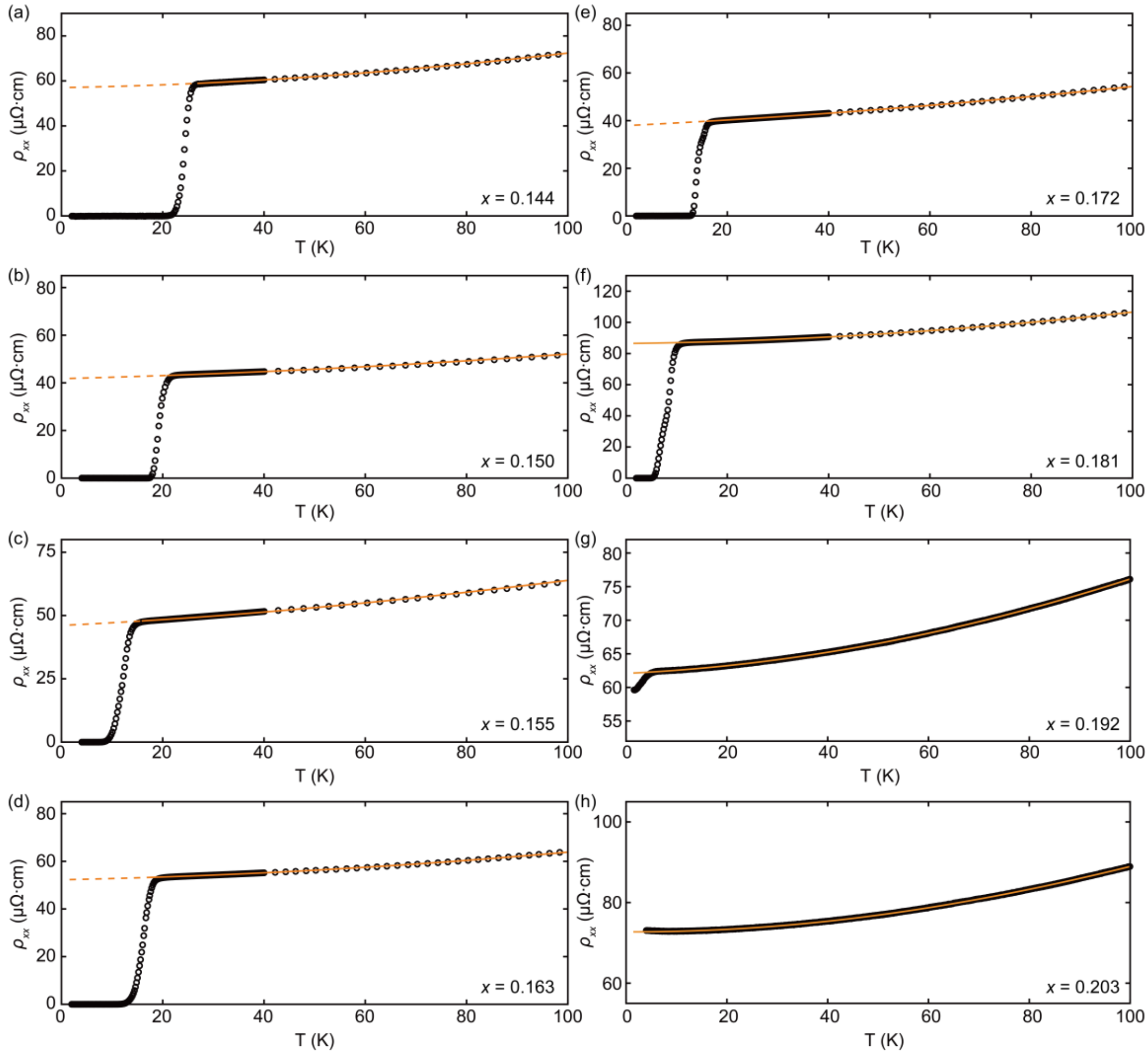

Fig. S8. (a-h) Temperature dependence of the resistivity for SNCO films with different doping levels, spanning from the optimal doping to the overdoped doping. The orange fitting curves show phenomenological fits of the experimental data using $\rho_{xx} = a_0 + a_1T + a_2T^2$.

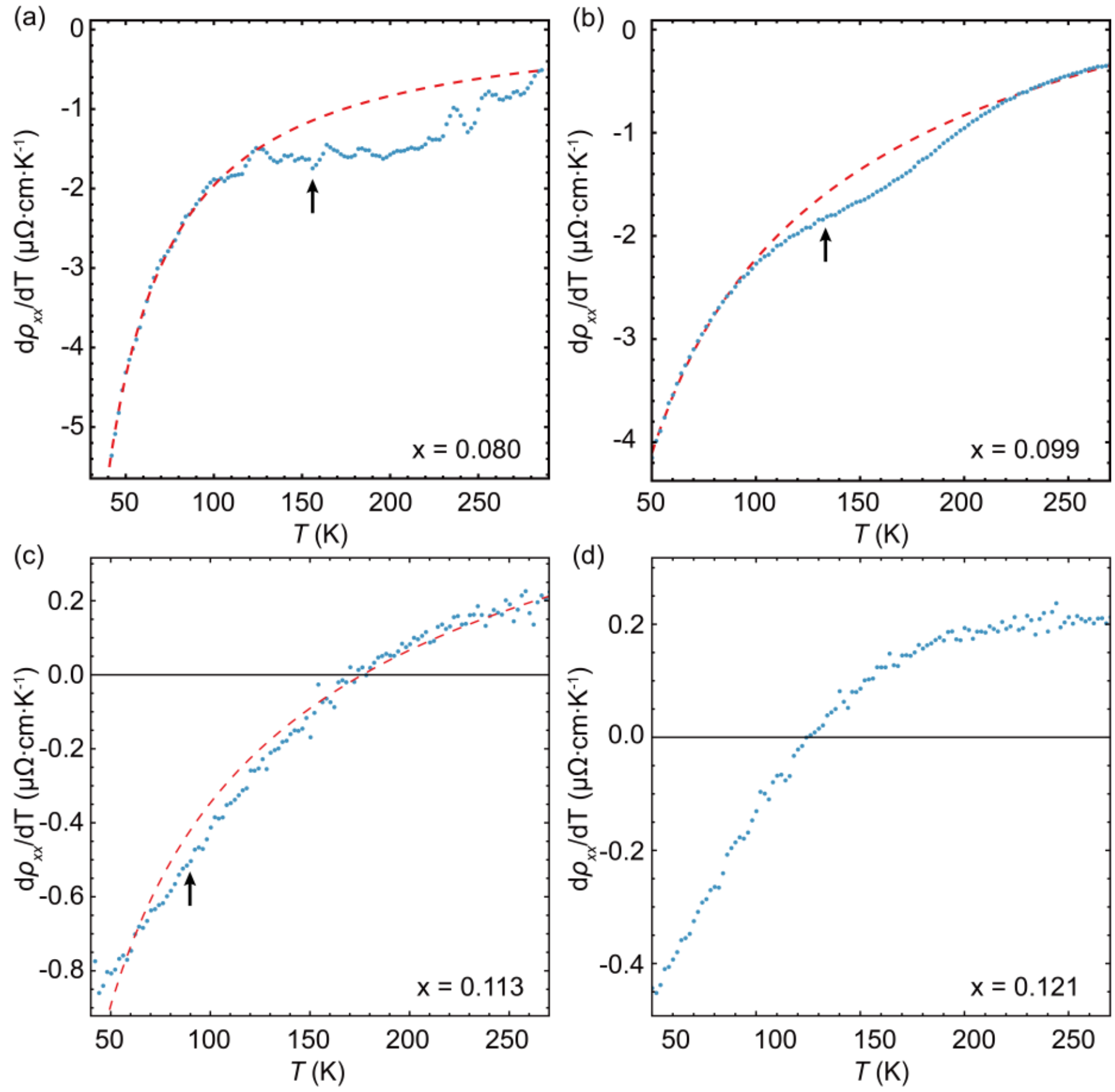

Fig. S9. The temperature dependence of the derivative of resistivity $d\rho_{xx}/dT$ and resistivity $\rho_{xx}(T)$ at Nd content (a) $x$ = 0.080, (b) $x$ = 0.99, (c) $x$ = 0.113, and (d) $x$ = 0.121. The red curves are phenomenological fits using $\frac{d\rho_{xx}}{dT} = a - \frac{b}{T^n}$. The resistivity hump in $\rho_{xx}(T)$ is magnified in the derivative plot. The black arrows correspond to the pink triangles in Fig. 1(d).